\documentclass[prb,showpacs,preprintnumbers,amsfonts,amssymb,amsmath,floats,twocolumn,aps,superscriptaddress]{revtex4}
 
\usepackage{amsmath,amssymb}
\usepackage{bm}
\usepackage{graphicx}
\usepackage{dcolumn}
\usepackage[final,dvips]{epsfig}

\newcommand{\be}{\begin{equation}}
\newcommand{\ee}{\end{equation}}

\newcommand{\ba}{\begin{eqnarray}}
\newcommand{\ea}{\end{eqnarray}}
\newcommand{\nn}{\nonumber \\}

\newcommand{\ff}[1]{{\bm #1}}
\newcommand{\Tr}{\mbox{Tr}}
\newcommand{\refeq}[1]{Eq.\ (\ref{eq:#1})}
\newcommand{\labeq}[1]{\label{eq:#1}}
\newcommand{\reffig}[1]{Fig.\ \ref{fig:#1}}

\newcommand{\rem}[1]{}

\begin{document}
 
 \title{Importance of local correlations for the order parameter of high-$T_c$ superconductors}
 
 \author{M.~Balzer}
 \affiliation{I. Institut f\"{ur} Theoretische Physik, Universit\"{a}t Hamburg, Germany }

 \author{W.~Hanke}
 \affiliation{Institut f\"{ur} Theoretische Physik und Astrophysik, Universit\"{a}t W\"{u}rzburg, Germany}

 \author{M.~Potthoff} 
 \affiliation{I. Institut f\"{ur} Theoretische Physik, Universit\"{a}t Hamburg, Germany }
   
\begin{abstract}
It is shown that local temporal correlations in addition to non-local spatial correlations are important to understand the size and the doping dependence of the d-wave superconducting order parameter of high-temperature superconductors.
To this end, the hole- and electron-doped two-dimensional Hubbard model at zero temperature is considered and treated by an extension of the variational cluster approximation. 
Within this approach, the effects of temporal correlations can be studied systematically and in a thermodynamically consistent way by comparing results obtained from different reference clusters. 
Contact can be made with previous cellular (plaquette) dynamical mean-field calculations. 
This shows that temporal correlations considerably decrease the order parameter and provide a substantial gain of binding energy.
Besides, a few methodical insights regarding real-space quantum-cluster approaches are obtained in addition.
\end{abstract}
 

 
\maketitle
 
\section{Introduction}
\label{sec1}

There are strong efforts devoted to a convincing theory of high-temperature superconductivity with non-local d-wave order parameter.
Some agreement has been achieved that essential physical properties can be studied within the prototypical two-dimensional single-band Hubbard model. \cite{And87}
Using standard notations the (grand-canonical) Hamiltonian reads:
\be
	H =  \sum_{ij\sigma} \left( t_{ij}-\mu \delta_{ij} \right)
	c^\dagger_{i\sigma} c_{j\sigma} 	+ U \sum_{i} n_{i\uparrow} n_{i\downarrow} \: ,
\labeq{ham}
\ee
with $c^\dagger_{i\sigma}$ ($c_{i\sigma}$) being the creation (annihilation) operator of an electron 
at lattice site $\bm r_i$ with spin projection $\sigma=\uparrow,\downarrow$.
$n_{i\sigma}=c^\dagger_{i\sigma} c_{i\sigma}$ denotes the corresponding occupation number
operator. 
The energy scale is set by the nearest-neighbor hopping $t_{nn}=-1$. 
To describe the universal low-energy physics of cuprates, an additional
second-nearest-neighbor hopping 
term with $t_{nnn} = 0.3$ \cite{ALJP95} as well as $U=8$ are taken as standard parameters
which will be kept fixed throughout the paper.
 
The physics of the intermediate- to strong-coupling regime of the model is notoriously complicated since there are several phases with different long- or short-range correlations competing on a low-energy scale. 
Due to the non-locality of the (d-wave) order parameter in the superconducting phase, a pure (single-site) mean-field approach cannot capture the essential features of the model. 
On the other hand, a direct numerical solution of finite but large two-dimensional Hubbard lattices is in principle possible via a quantum Monte-Carlo approach. \cite{Dag94,PHGE97}
However, for the interesting parameter regime, i.e.\ doped systems at low temperatures, the so-called sign problem makes simulations ineffective. 

In this situation, quantum-cluster approaches \cite{MJPH05,S08} appear promising, i.e.\ cluster extensions \cite{HTZ+98,LK00,KSPB01} of the dynamical mean-field theory (DMFT), \cite{GKKR96,KV04} or the variational cluster approximation (VCA). \cite{Pot03a,PAD03}
Common to all quantum-cluster approaches is that, for the calculation of the electronic self-energy, a small cluster is self-consistently or variationally embedded in a non-interacting bath that approximately accounts for the effects of the cluster environment.

Two types of correlations must be distinguished here:
(i) Spatial correlations are neglected completely in single-site DMFT but are included within a cluster approach, up to the size of the cluster. 
Here the number $L_c$ of correlated sites with finite Hubbard-$U$ in the cluster is essential.
For the doped two-dimensional Hubbard model, the important feedback of non-local magnetic correlations on the single-particle spectrum, for example, can only be captured by a cluster approach.
As spatial correlations are neglected beyond the size of the reference cluster, a quantum-cluster approach can also be seen as a cluster mean-field theory. 
(ii) Temporal correlations are fully accounted for already in single-site DMFT and give rise to a highly non-trivial frequency dependence of the self-energy which, for example, is vital to understand the Mott transition.
Here the number of uncorrelated sites with $U=0$ representing the bath is essential. 
As in the DMFT a continuum of bath degrees of freedom is used, it can be considered as an optimal mean-field theory. \cite{GKKR96,KV04,Pot03a}
If a continuum of bath sites is used, as it is intended usually, \cite{HTZ+98,LK00,KSPB01} an optimal quantum-cluster approach is generated for a given $L_c$.

Besides quantum Monte-Carlo techniques, the Lanczos method \cite{LG93} is frequently used as a {\em cluster solver} since this allows to study the ground-state phase diagram zero at temperature $T=0$.
As the Lanczos method is limited by the total number of sites, the variational cluster approximation (VCA), if combined with Lanczos as a solver, usually takes into account as many correlated sites as possible but completely disregards bath degrees of freedom. \cite{Pot03a,PAD03} 
The idea is that for a large cluster, temporal correlations are sufficiently accounted for since they are restored anyway in the infinite-cluster limit $L_c\to \infty$ where the VCA (as any quantum-cluster approach) becomes exact.

Using the VCA (no bath sites included), the ground-state phase diagram of the two-dimensional Hubbard model has been explored using clusters with up to $L_c=12$ sites and different cluster geometries. \cite{SLMT05,AA05,AAPH06a,AAPH06b,AAPH07}
Close to half-filling, for hole as well as for electron doping, these calculations reveal a phase where antiferromagnetic order and d-wave superconductivity are microscopically coexisting (AF+SC). 
With increasing doping a pure superconducting phase (SC) persists in both cases.
In agreement with experimental data, the phase diagram is asymmetric: 
Antiferromagnetic order, for example, extends to higher doping in the electron-doped case.
For the hole-doped system, an extended macroscopic phase separation of the mixed AF+SC phase with a purely superconduting phase SC can be found. \cite{AA05}
This has been interpreted as a tendency towards the formation of microscopically inhomogeneous (e.g.\ stripe) phases.
On the other hand, there is hardly a phase-separated state for the case of electron doping which may be seen to be consistent with the absence of rigorous signs for stripe structures in electron-doped materials. \cite{AA05,AAPH06b}

Using the cellular DMFT (including bath sites) for the set of hopping parameters considered here, \cite{KKS+08} but also for the particle-hole symmetric case ($t_{nnn}=0$), \cite{capone:054513} one can nicely reproduce the dome-like shape of the superconducting order parameter found in experiments. 
For $t_{nnn} = 0.3$ a homogeneous coexistence of antiferromagnetism and superconductivity (AF+SC) is found for fillings close to half-filling, while the pure superconducting phase (SC) persists for larger hole and electron concentrations. 
For the hole-doped system the pure superconducting phase extends over a broader range of dopings than on the electron-doped side.

We conclude that VCA (no bath sites) and C-DMFT (including bath sites) yield very similar ground-state phase diagrams. 
However, there is an obvious discrepancy with respect to the size of the superconducting order parameter.
VCA and C-DMFT results can differ by more than a factor 2 (see also results below).
In addition, the optimal (hole and electron) doping is larger in the VCA as compared to the C-DMFT calculations.
A major purpose of the present paper is to point out that this is due to an underestimation of local temporal correlations within conventional VCA.
Physically, this means that temporal correlations, giving rise e.g.\ to Kondo screening of magnetic moments, are important to understand the order parameter, the phase diagram and eventually the pairing mechanism. 

For practical calculations using the Lanczos method it is quite tempting to consider a plaquette of four correlated sites since this allows for d-wave order as well as non-local singlet formation with a minimum computational effort. 
Employing a plaquette represents the first important step beyond a single-site (dynamical) mean-field approach.
Here we present calculations obtained by an extension of the conventional VCA, i.e.\ we employ a reference cluster with one additional bath site attached to each correlated site.
This can be done within the framework of the SFT by comparing different but thermodynamically consistent approximations.

\section{Self-energy-functional theory}

\begin{figure}
	\centering
	\includegraphics[width=0.75\columnwidth,clip=]{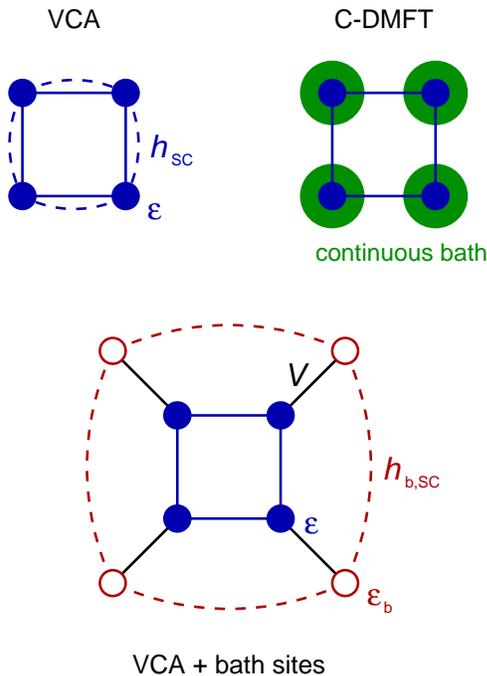}
	\caption{Reference systems and variational parameters as used for our calculations.
	Additionally a corresponding reference system generating the C-DMFT is illustrated.
Here, a continuous bath is optimized while the parameters of the correlated sites remain
at their physical values.
	}
	\label{fig:refsys}
\end{figure}

Both, the VCA and the C-DMFT, can be seen as approximations originating from a certain cluster reference system within the context set by the self-energy-functional theory (SFT). \cite{Pot03a,Pot05}
In the (conventional) VCA we consider a finite cluster without bath sites embedded in the original lattice while for C-DMFT a continuous bath is attached to each correlated site of a cluster (see \reffig{refsys}).
Both include non-local short-range spatial correlations on a scale up to the linear size of the cluster.
Beyond that both approximations are mean-field-like.
The treatment of local (temporal) correlations, however, is different:
The continuous bath considered in the C-DMFT ensures that these are taken into account exactly for arbitrary cluster size and even in the case of single-site DMFT.
On the other hand, within the VCA the description of local correlations becomes exact in the infinite cluster limit only.

Self-energy-functional theory \cite{Pot03a} starts from the grand potential of a system of interacting electrons expressed as a functional of the self-energy,
\be
	\Omega[\bm \Sigma] = \Tr \ln \left(\bm G^{-1}_0 - \bm \Sigma \right)^{-1} +	F[\bm\Sigma]
\: ,
	\labeq{SFT}	
\ee
with the free Green's function $\ff G_0$ and $F[\bm\Sigma]$ being the Legendre transform
of the Luttinger-Ward functional $\Phi[\ff G]$. \cite{LW60}
This functional can be shown to be stationary at the exact (physical) self-energy. 
Hence, we have the following dynamical variational principle:
\be
	\delta\Omega[\ff\Sigma] = 0 .
	\labeq{stationarity}
\ee
This, however, cannot be evaluated in practice since the functional form of $F[\bm\Sigma]$
(and of  $\Phi[\ff G]$) is actually unknown.
The main idea of the SFT is to restrict the variation of the self-energy in the
variational principle \refeq{stationarity} to a certain subspace of trial self-energies
which is spanned by the self-energies of an exactly solvable reference system (i.e.\ a
small cluster). 
This means to parameterize the trial self-energy $\ff\Sigma_{\ff t'}$ by the one-particle
parameters of the cluster $\ff t'$ and to treat $\ff t'$ as variational parameters: 
\be
	\frac{\partial}{\partial \ff t'}\Omega[\ff\Sigma_{\ff t'}] = 0 .
	\labeq{stationarity1}
\ee
For a small cluster, the value of the grand potential at the trial self-energy,
$\Omega[\ff\Sigma_{\ff t'}]$, and thus the condition \refeq{stationarity1} can be
evaluated numerically exact, see Refs.\ \onlinecite{Pot03a,AAPH06b} for examples.

For any practical calculation, one usually considers a few physically relevant parameters
only.
\reffig{refsys} illustrates the different plaquette reference systems considered in this
study and, for the VCA, the one-particle parameters to be optimized. 
The conventional VCA refers to a reference system with the following Hamiltonian:
\ba
	H' & = & 	\sum_{ij\sigma} t'_{ij} c^\dagger_{i\sigma} c_{j\sigma} 
		+ \left(\varepsilon-\mu\right) \sum_{i\sigma} n_{i\sigma}
		+ U \sum_i n_{i\uparrow} n_{i\downarrow} \nn
	& &	\mbox{}+ h_{\rm SC} \sum_{ij} \Delta_{ij} \left( c_{i\downarrow} c_{j\uparrow} +
				\mbox{H.c.} \right)
\label{eq:RefSC0}
\ea
including a global shift of the on-site energies $\varepsilon$ and a ficticious
symmetry-breaking Weiss field of strength $h_{\rm SC}$ as variational parameters. 
Optimization of $\varepsilon$ ensures thermodynamic consistency with respect to the total
particle number
\cite{AAPH06a} while the Weiss field allows for a possible superconducting phase.
Note that the interaction term is the same as for the original model \refeq{ham}.
Contrary, only intra-cluster hopping parameters $t'_{ij}$ are retained while the
inter-cluster hopping is switched off.
For $d_{{x^2}\mathrm{-}{y^2}}$-pairing,
\be
	\Delta_{ij} = 
		\left\{\begin{array}{rcc}
			1 & \mathrm{for} & \bm r_i - \bm r_j = \pm \bm e_x \\
			-1 & \mathrm{for} & \bm r_i - \bm r_j = \pm \bm e_y
		\end{array}	\right. 
\ee
with cluster sites $\bm r_i$, $\bm r_j$ and $\bm e_{x}$, $\bm e_{y}$ being the unity
vectors in the $x$ and $y$ direction.

Attaching one additional bath site to each correlated site results in the reference system
\reffig{refsys}, bottom. 
The Hamiltonian reads:
\ba
	H' & = & 	\sum_{ij\sigma}  t'_{ij}c^\dagger_{i\sigma} c_{j\sigma} +
			\left(\varepsilon-\mu\right)	\sum_{i\sigma} n_{i\sigma} 
				 + U \sum_i n_{i\uparrow} n_{i\downarrow}		 \nn
	& & \mbox{} + V \sum_{i\sigma} \left( a^\dagger_{i\sigma} c_{i\sigma} + \mbox{H.c.}
			\right)
			+  \left(\varepsilon_b-\mu\right) \sum_{i\sigma} n_{{\rm b,}i\sigma} \nn
	& &	\mbox{}+ h_{\rm b,SC} \sum_{ij} \Delta_{ij} \left( a_{i\downarrow} a_{j\uparrow} +
				\mbox{H.c.} \right) .
\label{eq:RefSC}
\ea
Here, the hybridization strength $V$ with the bath sites and their on-site energy
$\varepsilon_b$ are treated as additional variational parameters.
This leads to an improved description of local correlations as compared to the reference
system \refeq{RefSC0}.
In \refeq{RefSC} $a^\dagger_{i\sigma}$ ($a_{i\sigma}$) creates (annihilates) an electron
with spin projection
$\sigma=\uparrow,\downarrow$ on a bath site coupled via $V$ to the cluster site $\bm r_i$,
and $n_{{\rm b},i\sigma} = a^\dagger_{i\sigma} a_{i\sigma}$.

Within the SFT and using the reference system in \reffig{refsys} (right), it turns out
that the Euler equation of the variational principle \refeq{stationarity1} is equivalent
with the C-DMFT self-consistency equation for the parameters of the continuous bath. 
The optimal values of the one-particle parameters associated with correlated sites can be
shown to be given by their physical values, i.e.\ those of the original system.
\cite{AAPH06a}
For a reference system with a finite number of bath sites, on the other hand, this is no
longer necessarily true.
Namely, the truncation of the inter-cluster hopping is partially compensated for by
optimal parameters associated with the correlated sites that differ from their physical
values.
In the conventional VCA (i.e.\ no bath sites) this is obvious.

For a refererence system including bath sites, \reffig{refsys} (bottom), we therefore
expect that applying the ficticious (i.e.\ unphysical) Weiss field to the bath sites is
much more efficient as compared to the correlated sites.
We have numerically checked this for the case of antiferromagnetic order where both, a
staggered magnetic Weiss field on the correlated and on the bath sites have been varied
simultaneously: 
The optimal staggered field on the correlated sites has turned out to be typically more
than an order of magnitude smaller than the one on the bath sites and to be negligible for
the calculation of observables.

Consequently, in \refeq{RefSC} we attach the symmetry breaking Weiss field $h_{\rm b,SC}$
to the
bath sites.
Note, however, that we still do optimize the parameter $\varepsilon$ (in addition to
$\varepsilon_b$) to ensure the above-mentioned consistency with respect to the particle
number.

\section{Spin-dependent particle-hole transformation}

As has been demonstrated by S\'en\'echal et al.\ \cite{SLMT05} a possible way to treat
superconductivity within VCA is to employ the Nambu formalism where normal as well as
anomalous Green's functions have to be computed to evaluate the self-energy functional.
An alternative is given by a spin-dependent particle-hole transformation of the original
and the reference system. 
This restores particle-number conservation and avoids anomalous Green's functions. 
The transformation is given by
\ba
	c^\dagger_{i\uparrow} \to d^\dagger_{i\uparrow} & \hspace{3em}
	a^\dagger_{i\uparrow} \to b^\dagger_{i\uparrow} \nn
	c^\dagger_{i\downarrow} \to d_{i\downarrow} &  \hspace{3em}
	a^\dagger_{i\downarrow} \to b_{i\downarrow} \; .
\label{eq:SCTLT}
\ea
Applying the particle-hole transformation to $H'$, \refeq{RefSC}, we get
\ba
	\widetilde{H}' & = &
	 \sum_{ij} t'_{ij} \left( d^\dagger_{i\uparrow} d_{j\uparrow}
		 - d^\dagger_{i\downarrow}	d_{j\downarrow} \right) \nn
		& & \mbox{}+ \left(\varepsilon-\mu\right) \sum_i \left(\widetilde{n}_{i\uparrow} -
		\widetilde{n}_{i\downarrow}\right) 		+ U \sum_i \widetilde{n}_{i\uparrow} \nn
		& & \mbox{}- U \sum_i \widetilde{n}_{i\uparrow} \widetilde{n}_{i\downarrow} \nn
		& & \mbox{}+ V \sum_i \left( b^\dagger_{i\uparrow} d_{j\uparrow}
		 - b^\dagger_{i\downarrow}	d_{j\downarrow} \right) \nn
		& & \mbox{}+ \left(\varepsilon_b-\mu\right) \sum_i \left(\widetilde{n}_{{\rm
		b,}i\uparrow} -
\widetilde{n}_{{\rm b,}i\downarrow}\right)  \nn
		& & + h_{\rm b,SC} \sum_{ij} \Delta_{ij} \left( b^\dagger_{i\uparrow} b_{j\downarrow}
		+ b^\dagger_{i\downarrow} b_{j\uparrow} \right) \nn
	 & &\mbox{} + \; \sum_i \left( \varepsilon + \varepsilon_b - 2\mu \right) \; ,
\label{eq:SCtrans}
\ea
where $\widetilde{n}_{i\sigma} = d^\dagger_{i\sigma} d_{i\sigma}$, 
$\widetilde{n}_{{\rm b,}i\sigma} = b^\dagger_{i\sigma} b_{i\sigma}$ and $t'_{ij} =
t'_{ji}$ and $\Delta_{ij}=\Delta_{ji}$ have been used.

\refeq{SCtrans} is interpreted as a reference system corresponding to the particle-hole
transformed original system \refeq{ham}.
For both cases, the transformation yields an attractive Hubbard interaction ($U \to -U$). 
Hopping terms become spin-dependent, and chemical-potential-like terms are mapped onto
ferromagnetic fields and vice versa. 
The nonlocal pairing field maps onto a nonlocal spin-flip term.
The transformed Hamiltonian respects particle-number conservation (unless there were
spin-flip terms in the original one which would transform to pairing fields), the
$z$-component of the total spin, however, is no longer conserved.
For the transformed reference system, the anomalous Green's function vanishes; the normal
Green's function is no longer diagonal in the spin index $\sigma$. 

In practice, we first evaluate the self-energy functional for the transformed original
system $\widetilde{H}$ by solving the problem posed by the transformed reference system
$\widetilde{H}'$, i.e.\ we calculate the corresponding Green's function and self-energy,
evaluate therewith the self-energy functional and optimize the variational parameters.
Note that the variational parameters are chosen accordingly (e.g.\ the on-site energy is
represented as the strength of a ferromagnetic field in the transformed system, see
\refeq{SCtrans}, and this field strength is varied and optimized).
We have crosschecked with the results obtained without transformation where possible.
Observables, i.e.\ static expectation values, Green's functions and the SFT grand
potential, have to be transformed back, optimization of the parameters is performed
subsequently. 

As an advantage of using the particle-hole transformation, only minor changes of the
standard VCA code are necessary. The evaluation of the self-energy functional can still be
done using the Q-matrix technique introduced in Ref.\ \onlinecite{AAPH06b}, for example.

\section{Results and Discussion}

For the discussion of the results of our VCA calculations we focus on the superconducting order parameter. 
Besides integral quantities like the ground-state energy, this is the observable which is sensitive  to the relevant low-energy scale and which shows the largest discrepancy when comparing (plain) VCA and C-DMFT. 

\begin{figure}[t]
	\centering
	\includegraphics[width=0.99\columnwidth,clip=]{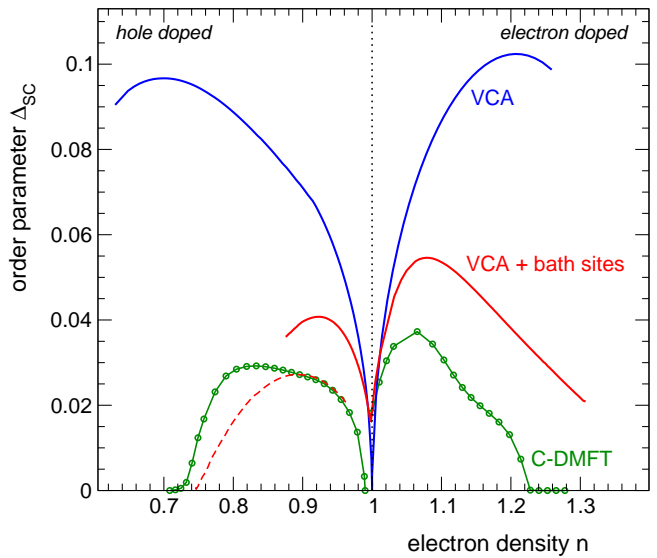}
	\caption{Superconducting (d-wave) order parameter $\Delta_\mathrm{SC}$ (see
\refeq{DefDeltaSC}) as a function of the electron density $n$ within conventional VCA
(blue line, variational parameters
	$\varepsilon$ and $h_{\rm SC}$) and VCA including bath sites (red line, variational
parameters 
	$\varepsilon$, $\varepsilon_b$, $V$ and $h_\mathrm{b,SC}$). In the case of VCA
	with bath sites a second solution with higher ground-state energy is found (dashed red
line). 
	Calculations have been performed for system sizes up to $L=6400$ sites.
	C-DMFT results from Kancharla et al. (green symbols, Ref.\ \onlinecite{KKS+08}, Fig. 2) are
shown for comparison.}
	\label{fig:DeltaNsup}
\end{figure}

To be consistent with the  C-DMFT calculations of Kancharla et al.\ \cite{KKS+08} we take 
\be
	\Delta_\mathrm{SC} = \left|\langle{c_{i\downarrow} c_{j\uparrow}}\rangle\right|
\label{eq:DefDeltaSC}	
\ee
as the definition for the superconducting order parameter.
The indices $i$ and $j$ refer to nearest-neighbor sites.
Since translational symmetry is broken by the cluster approach, 
$\langle{c_{i\downarrow} c_{j\uparrow}}\rangle$ is computed for
lattice sites which, in the reference system, would belong to the same cluster.
VCA calculations using the reference systems given in \reffig{refsys} have been performed
for a square lattice consisting of up to $L=6400$ sites for nearest-neighbor hopping set
to $t_{nn}=-1$, second-nearest neighbor hopping $t_{nnn} = 0.3$ and $U=8$.

The results are summarized in \reffig{DeltaNsup} which shows the order parameter
$\Delta_{\rm SC}$ as a function of the electron density $n$ for zero temperature. 
Apart from the normal state, we have allowed for a pure superconducting phase only and
suppressed a possible 
antiferromagnetic phase which shows up close to half-filling and has been studied in
detail in previous work. \cite{DAH+04,AA05,AAPH06a,AAPH06b,AAPH07}
In the case of the conventional VCA (blue line, reference system \reffig{refsys}, left), a
Mott insulating solution is found at half-filling. 
With increasing doping, $\Delta_{\rm SC}$ increases and reveals its maximum value at $n
\approx 0.7$ for hole doping and $n \approx 1.2$ for electron doping. 
Unfortunately, we could not trace the solution for dopings much larger than the optimal
dopings. 
Quite generally, a possible discontinuous change of the ground state of the finite
reference cluster may result in a discontinuous change of the self-energy and thus of the
SFT grand potential. 
Thereby, a solution can cease to exist.

Using VCA with bath sites (red line, reference system \reffig{refsys}, bottom) the order
parameter is small but remains finite at half-filling. 
The maxima of $\Delta_{\rm SC}$ are found at $n\approx 0.92$ and $n \approx 1.08$, i.e.\
at significantly lower dopings as compared to the results of the conventional VCA.
This comes close to the C-DMFT results of Kancharla et al.\ \cite{KKS+08}
The stationary point could not be traced for hole doping larger than $1-n \approx 0.2$ and
electron doping larger than $n-1 \approx 0.3$.
On the hole-doped side, a second symmetry-broken solution could be found which comes very
close to the C-DMFT data and shows up a second-order critical point with $\Delta_{\rm SC}
\to 0$ and with the symmetry-breaking Weiss field $h_{\rm b,SC} \to 0$ at $n\approx 0.75$.
This solution is found to exist up to $n \lesssim 0.96$.

\begin{figure}
	\centering
	\includegraphics[width=0.99\columnwidth,clip=]{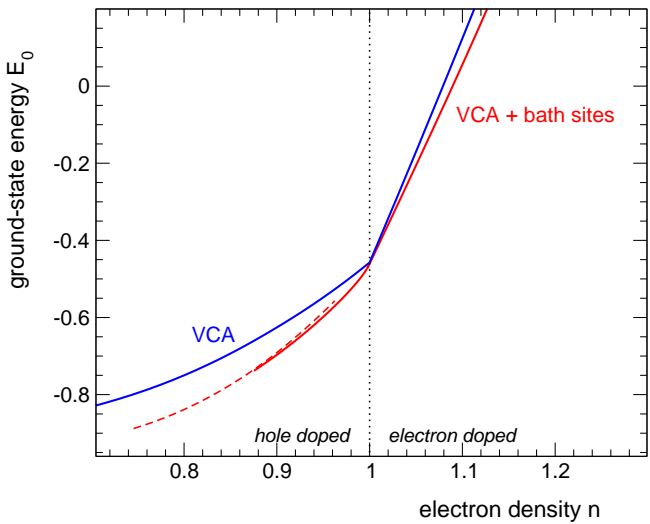}
	\caption{Ground-state energy as a function of the electron density $n$ within
conventional VCA (blue line) and VCA including bath sites (red line). Dashed red line:
second solution with higher ground-state energy.}
	\label{fig:energy}
\end{figure}

In the entire hole-doping range where two superconducting solutions can be found, however,
the one with the larger order parameter is more stable, i.e.\ for $T=0$ has a lower
ground-state energy at the same filling.
This is what could be expected from a (cluster) mean-field theory. 
The ground-state energies of the respective solutions are shown in \reffig{energy}. 
Note that the energy difference between the two solutions obtained from VCA with bath
sites is rather small compared to the energy difference between the results from
conventional VCA and VCA with bath sites. 
It is also worth mentioning that the conventional VCA ground-state energy is higher
although the corresponding solution has the larger order parameter.
Both facts show that including bath sites considerably improves the approximation; bath
sites couple to the plaquette of correletad sites and yield a significant binding-energy
gain.
This gain is finite but small at half-filling and increases with increasing doping.

Conceptually, the inclusion of bath sites improves the variational ansatz for the trial
self-energy.
Bath sites mimic the residual lattice into which the cluster is embedded by taking into
account processes involving the sites of the cluster environment on a mean-field level.
Physically, this means to improve the description of local (intra-cluster) quantum
fluctuations.
These additional fluctuations are expected and in fact seen to decrease the order
parameter. 
An in principle optimal treatment of local fluctuations is provided by (cluster) DMFT
where a continuum of bath sites is considered. 
By comparing the VCA with and without bath sites with the C-DMFT and the respective
results for the order parameter (see \reffig{DeltaNsup}), we find that the essential step
is already done by attaching a {\em single} bath site to each of the correlated sites.
The convergence with the number of bath sites has already been recognized to be extremely
fast in different contexts, see Refs.\ \onlinecite{Poz04,BHP08,BKS+09}.
The inclusion of more and more bath sites must brigde the remaining discrepancies with the
C-DMFT results since conceptually the SFT recovers C-DMFT in this limit. 
Note, however, that for the practical C-DMFT calculations performed at $T=0$ using the
Lanczos technique, a plaquette geometry with 2 additional bath sites per correlated site
has been used only. \cite{KKS+08}

\begin{figure}
	\centering
	\includegraphics[width=0.99\columnwidth,clip=]{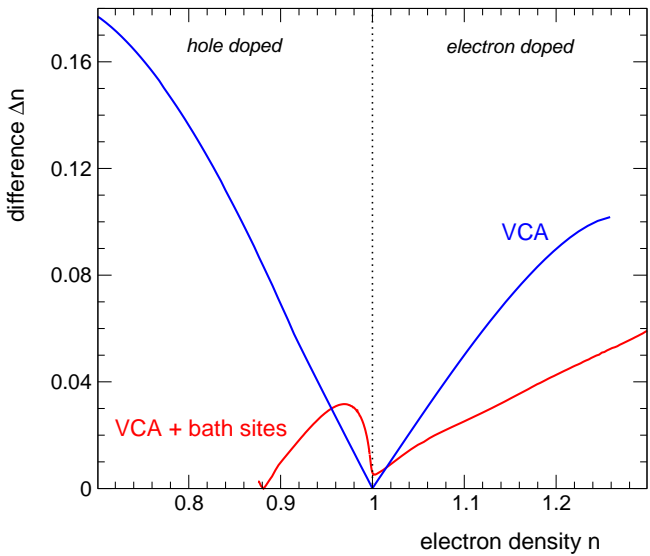}
	\caption{Difference $|n-n'|$ of the electron density
		in the original and the reference system (correlated sites only)
		as a function of the electron density $n$. VCA calculations with and without bath sites
                corresponding to the calculations shown in \reffig{DeltaNsup}.  
		$n-n'$ changes sign at $n=1$ for the conventional VCA calculation (blue
		line) and is positive for electron doping.
	}
	\label{fig:SCel}
\end{figure}

For the conventional VCA without bath sites, the normal state at half-filling $n=1$ is
described by a half-filled reference cluster, i.e.\ $n'=1$.
Upon doping and provided that the solution can be traced continuously, i.e.\ that there is
no discontinuous change of the cluster ground state, the cluster remains at half-filling,
$n'=1$.
This yields an obviously unphysical description of strongly doped phases. 
As has already been demonstrated for the one-dimensional Hubbard model, \cite{BHP08} bath
sites can help to overcome this problem since they serve as charge reservoirs:
While the reference cluster remains half-filled, the average occupation number $n'$ on the
correlated sites in the cluster is close to the electron density $n$ of the lattice
model. 
For a superconducting state the situation is somewhat different since there is no
particle-number conservation. 
Still there is a similar problem for plain VCA calculations without bath sites as can be
seen from \reffig{SCel}.
In the hole-doped case at $n=0.7$, for example, we find $n' \approx 0.88$ (this
corresponds to a sizeable difference $|n-n'| \approx 0.18$, see figure).
Adding a single bath site per correlated site, yields a strongly improved although not
perfect description with a smaller difference $|n-n'|$ as shown in the figure.
Conceptually, $n=n'$ can be achieved for a continuum of bath sites only, i.e.\ for full
C-DMFT.

As can be seen in \reffig{SCel} there is a small but finite difference $|n-n'|$ even at half-filling opposed to the results of the conventional VCA. 
This might correspond to the finite but small value of the order parameter at half-filling (see \reffig{DeltaNsup}) which then would have to be considered as an artifact. 
On the other hand, a superconducting state at half-filling is not unexpected for frustrated lattice models (i.e.\ for finite $t_{nnn}$) 
if, at the critical point for the superconducting instability, the system is metallic.
In fact, a $d$-wave superconducting state at $n=1$ has been found, for example, even within plain VCA in Ref.\ \onlinecite{AAPH07} and within variational Monte Carlo. \cite{YOT06}

\begin{figure}
	\centering
	\includegraphics[width=0.99\columnwidth,clip=]{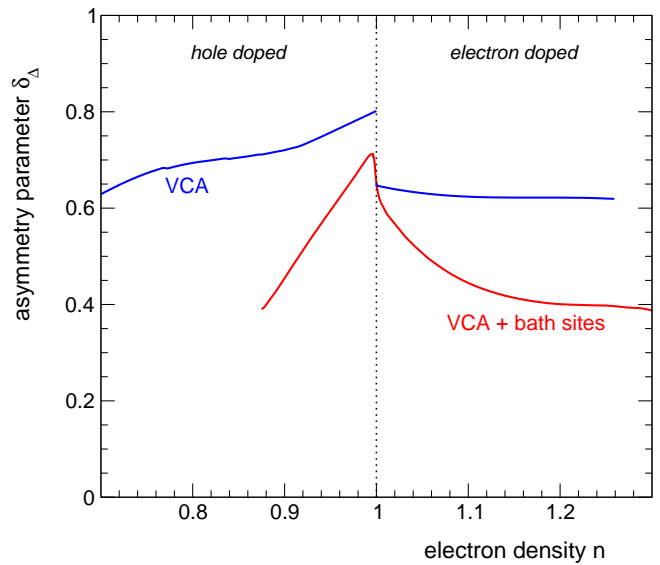}
	\caption{
	Relative difference $\delta_\Delta$ according to \refeq{relDelta} as function of $n$ for
	VCA calculations with and without bath sites. We find the bath sites to partially
compensate
	for the artificial breaking of translational symmetry.
	}
	\label{fig:DeltaSC}
\end{figure}

The translational symmetries of the original lattice are artificially broken by any
(real-space) cluster approximation. 
Since bath sites are expected to provide an improved embedding of the cluster in the
residual lattice, however, one may expect that they help to restore the symmetries of the
original state in a quantum-cluster approach.
It is clear that even with a full C-DMFT calculation this can be achieved only partially
unless one uses additional (ad hoc) symmetrization schemes subsequently.
Nevertheless, there is a considerable improvement as shown by our results.
In \refeq{DefDeltaSC} we have defined the order parameter $\Delta_{\rm SC}$ on neighboring
sites in the original lattice belonging to the same reference cluster. 
$\Delta_{\rm SC}$ is independent of the nearest-neighbor pair for a $2\times 2$ cluster.
To compare with the order parameter $\Delta_{\rm SC, inter}$ for neighboring sites
belonging to different clusters, we define the asymmetry parameter
\be
	\delta_\Delta = \frac{\Delta_{\rm SC} - \Delta_{\rm SC, inter}}
	{\Delta_{\rm SC} + \Delta_{\rm SC, inter}}
	\labeq{relDelta} \;.
\ee
\reffig{DeltaSC} shows $\delta_\Delta$ as a function of the electron density $n$. 
It is remarkable that the asymmetry can be reduced, depending on the doping, by more than
one third.
We expect that for an even stronger reduction a larger cluster would be much more
efficient than adding more bath sites.

\section{Conclusion}

Several cluster-embedding approaches can be formulated within the framework of the
self-energy-functional theory which differ with respect to bath degrees of freedom. 
Cellular DMFT maps the lattice problem onto a small cluster of correlated sites to each of
which a continuous set of bath sites is attached.
This ensures an optimal description of the local fluctuations. 
To access the zero-temperature phase diagram for Hubbard-type models of strongly
correlated electron systems, there is basically the Lanczos technique only which may serve
as a ``solver'' for the effective impurity (cluster) problem. 
This implies that in practice a limited number of bath sites can be taken into account. 
The plain variational cluster approach (VCA) employs a reference system without any bath
sites at all. 
This approximation represents a valuable counterpart to the C-DMFT which, however,
provides an exact treatment of the local fluctuations in the limit of an infinitely large
reference cluster only.

An extended VCA which employs a reference system with a single bath site per correlated site suggests itself as a compromise and has been used here to study the doped two-dimensional Hubbard model at zero temperature. 
As compared to conventional (plain) VCA, we found a substantial gain in binding energy when attaching the bath sites. 
For a pure d-wave superconducting state, the extended VCA yields a considerably smaller order parameter which comes close to the predictions of plaquette C-DMFT. 
The same holds for the optimal values for hole and electron doping, defined as maxima of the filling-dependent order parameter.
From the comparison with the C-DMFT data we conclude that, as concerns the improved description of local correlations, attaching a {\em single} bath site per correlated site does the main part of the job. 
Clearly, larger clusters exceeding a plaquette are desirable to improve the description of spatial correlations.

Methodically, bath sites serve as charge reservoirs and in this way yield an average occupation on the correlated sites in the cluster which, opposed to plain VCA, is much closer to the band filling in the original lattice model although the total cluster filling still stays at half-filling in the normal state (this is different for the superconducting phase where the particle number has no definite value).
Bath sites also help to partially restore the translational symmetry that is artificially broken by the (real-space) quantum-cluster concept.

Physically, the comparison of the different variational cluster approximations shows that local temporal correlations in addition to non-local spatial correlations are important to understand the size and the doping dependence of the d-wave superconducting order parameter. 
The corrections due to local correlations are most pronounced for the optimal-doped and the overdoped regime. 
Here, local quantum fluctuations are found to substantially decrease the order parameter. 
This could be attributed to a Kondo-type screening of local magnetic moments which in turn reduces spin fluctuations.
\\

We would like to thank M. Aichhorn for discussions.
Support of this work by the Deutsche Forschungsgemeinschaft (Forschergruppe FOR 538/P1) and
by the Landesexzellenzinitiative Hamburg is gratefully acknowledged.
\\


\begin{thebibliography}{28}
\expandafter\ifx\csname natexlab\endcsname\relax\def\natexlab#1{#1}\fi
\expandafter\ifx\csname bibnamefont\endcsname\relax
  \def\bibnamefont#1{#1}\fi
\expandafter\ifx\csname bibfnamefont\endcsname\relax
  \def\bibfnamefont#1{#1}\fi
\expandafter\ifx\csname citenamefont\endcsname\relax
  \def\citenamefont#1{#1}\fi
\expandafter\ifx\csname url\endcsname\relax
  \def\url#1{\texttt{#1}}\fi
\expandafter\ifx\csname urlprefix\endcsname\relax\def\urlprefix{URL }\fi
\providecommand{\bibinfo}[2]{#2}
\providecommand{\eprint}[2][]{\url{#2}}

\bibitem[{\citenamefont{Anderson}(1987)}]{And87}
\bibinfo{author}{\bibfnamefont{P.~W.} \bibnamefont{Anderson}},
  \bibinfo{journal}{Science} \textbf{\bibinfo{volume}{235}},
  \bibinfo{pages}{1196} (\bibinfo{year}{1987}).

\bibitem[{\citenamefont{Andersen et~al.}(1995)\citenamefont{Andersen,
  Liechtenstein, Jepsen, and Paulsen}}]{ALJP95}
\bibinfo{author}{\bibfnamefont{O.~K.} \bibnamefont{Andersen}},
  \bibinfo{author}{\bibfnamefont{A.~I.} \bibnamefont{Liechtenstein}},
  \bibinfo{author}{\bibfnamefont{O.}~\bibnamefont{Jepsen}}, \bibnamefont{and}
  \bibinfo{author}{\bibfnamefont{F.}~\bibnamefont{Paulsen}},
  \bibinfo{journal}{J. Phys. Chem. Solids} \textbf{\bibinfo{volume}{56}},
  \bibinfo{pages}{1573} (\bibinfo{year}{1995}).

\bibitem[{\citenamefont{Dagotto}(1994)}]{Dag94}
\bibinfo{author}{\bibfnamefont{E.}~\bibnamefont{Dagotto}},
  \bibinfo{journal}{Rev. Mod. Phys.} \textbf{\bibinfo{volume}{66}},
  \bibinfo{pages}{763} (\bibinfo{year}{1994}).

\bibitem[{\citenamefont{Preuss et~al.}(1997)\citenamefont{Preuss, Hanke,
  Gr\"ober, and Evertz}}]{PHGE97}
\bibinfo{author}{\bibfnamefont{R.}~\bibnamefont{Preuss}},
  \bibinfo{author}{\bibfnamefont{W.}~\bibnamefont{Hanke}},
  \bibinfo{author}{\bibfnamefont{C.}~\bibnamefont{Gr\"ober}}, \bibnamefont{and}
  \bibinfo{author}{\bibfnamefont{H.~G.} \bibnamefont{Evertz}},
  \bibinfo{journal}{Phys. Rev. Lett.} \textbf{\bibinfo{volume}{79}},
  \bibinfo{pages}{1122} (\bibinfo{year}{1997}).

\bibitem[{\citenamefont{Maier et~al.}(2005)\citenamefont{Maier, Jarrell,
  Pruschke, and Hettler}}]{MJPH05}
\bibinfo{author}{\bibfnamefont{T.}~\bibnamefont{Maier}},
  \bibinfo{author}{\bibfnamefont{M.}~\bibnamefont{Jarrell}},
  \bibinfo{author}{\bibfnamefont{T.}~\bibnamefont{Pruschke}}, \bibnamefont{and}
  \bibinfo{author}{\bibfnamefont{M.~H.} \bibnamefont{Hettler}},
  \bibinfo{journal}{Rev. Mod. Phys.} \textbf{\bibinfo{volume}{77}},
  \bibinfo{pages}{1027} (\bibinfo{year}{2005}).

\bibitem[{\citenamefont{S\'en\'echal}(2008)}]{S08}
\bibinfo{author}{\bibfnamefont{D.}~\bibnamefont{S\'en\'echal}},
  \bibinfo{journal}{preprint cond-mat} p. \bibinfo{pages}{arXiv:0806.2690v1}
  (\bibinfo{year}{2008}).

\bibitem[{\citenamefont{Hettler et~al.}(1998)\citenamefont{Hettler,
  Tahvildar-Zadeh, Jarrell, Pruschke, and Krishnamurthy}}]{HTZ+98}
\bibinfo{author}{\bibfnamefont{M.~H.} \bibnamefont{Hettler}},
  \bibinfo{author}{\bibfnamefont{A.~N.} \bibnamefont{Tahvildar-Zadeh}},
  \bibinfo{author}{\bibfnamefont{M.}~\bibnamefont{Jarrell}},
  \bibinfo{author}{\bibfnamefont{T.}~\bibnamefont{Pruschke}}, \bibnamefont{and}
  \bibinfo{author}{\bibfnamefont{H.~R.} \bibnamefont{Krishnamurthy}},
  \bibinfo{journal}{Phys. Rev. B} \textbf{\bibinfo{volume}{58}},
  \bibinfo{pages}{R7475} (\bibinfo{year}{1998}).

\bibitem[{\citenamefont{Lichtenstein and Katsnelson}(2000)}]{LK00}
\bibinfo{author}{\bibfnamefont{A.~I.} \bibnamefont{Lichtenstein}}
  \bibnamefont{and} \bibinfo{author}{\bibfnamefont{M.~I.}
  \bibnamefont{Katsnelson}}, \bibinfo{journal}{Phys. Rev. B}
  \textbf{\bibinfo{volume}{62}}, \bibinfo{pages}{R9283} (\bibinfo{year}{2000}).

\bibitem[{\citenamefont{Kotliar et~al.}(2001)\citenamefont{Kotliar, Savrasov,
  P\'alsson, and Biroli}}]{KSPB01}
\bibinfo{author}{\bibfnamefont{G.}~\bibnamefont{Kotliar}},
  \bibinfo{author}{\bibfnamefont{S.~Y.} \bibnamefont{Savrasov}},
  \bibinfo{author}{\bibfnamefont{G.}~\bibnamefont{P\'alsson}},
  \bibnamefont{and} \bibinfo{author}{\bibfnamefont{G.}~\bibnamefont{Biroli}},
  \bibinfo{journal}{Phys. Rev. Lett.} \textbf{\bibinfo{volume}{87}},
  \bibinfo{pages}{186401} (\bibinfo{year}{2001}).

\bibitem[{\citenamefont{Georges et~al.}(1996)\citenamefont{Georges, Kotliar,
  Krauth, and Rozenberg}}]{GKKR96}
\bibinfo{author}{\bibfnamefont{A.}~\bibnamefont{Georges}},
  \bibinfo{author}{\bibfnamefont{G.}~\bibnamefont{Kotliar}},
  \bibinfo{author}{\bibfnamefont{W.}~\bibnamefont{Krauth}}, \bibnamefont{and}
  \bibinfo{author}{\bibfnamefont{M.~J.} \bibnamefont{Rozenberg}},
  \bibinfo{journal}{Rev. Mod. Phys.} \textbf{\bibinfo{volume}{68}},
  \bibinfo{pages}{13} (\bibinfo{year}{1996}).

\bibitem[{\citenamefont{Kotliar and Vollhardt}(2004)}]{KV04}
\bibinfo{author}{\bibfnamefont{G.}~\bibnamefont{Kotliar}} \bibnamefont{and}
  \bibinfo{author}{\bibfnamefont{D.}~\bibnamefont{Vollhardt}},
  \bibinfo{journal}{Physics Today} \textbf{\bibinfo{volume}{57}},
  \bibinfo{pages}{53} (\bibinfo{year}{2004}).

\bibitem[{\citenamefont{Potthoff}(2003)}]{Pot03a}
\bibinfo{author}{\bibfnamefont{M.}~\bibnamefont{Potthoff}},
  \bibinfo{journal}{Euro. Phys. J. B} \textbf{\bibinfo{volume}{32}},
  \bibinfo{pages}{429} (\bibinfo{year}{2003}).

\bibitem[{\citenamefont{Potthoff et~al.}(2003)\citenamefont{Potthoff, Aichhorn,
  and Dahnken}}]{PAD03}
\bibinfo{author}{\bibfnamefont{M.}~\bibnamefont{Potthoff}},
  \bibinfo{author}{\bibfnamefont{M.}~\bibnamefont{Aichhorn}}, \bibnamefont{and}
  \bibinfo{author}{\bibfnamefont{C.}~\bibnamefont{Dahnken}},
  \bibinfo{journal}{Phys. Rev. Lett.} \textbf{\bibinfo{volume}{91}},
  \bibinfo{pages}{206402} (\bibinfo{year}{2003}).

\bibitem[{\citenamefont{Lin and Gubernatis}(1993)}]{LG93}
\bibinfo{author}{\bibfnamefont{H.~Q.} \bibnamefont{Lin}} \bibnamefont{and}
  \bibinfo{author}{\bibfnamefont{J.~E.} \bibnamefont{Gubernatis}},
  \bibinfo{journal}{Comput. Phys.} \textbf{\bibinfo{volume}{7}},
  \bibinfo{pages}{400} (\bibinfo{year}{1993}).

\bibitem[{\citenamefont{S\'en\'echal et~al.}(2005)\citenamefont{S\'en\'echal,
  Lavertu, Marois, and Tremblay}}]{SLMT05}
\bibinfo{author}{\bibfnamefont{D.}~\bibnamefont{S\'en\'echal}},
  \bibinfo{author}{\bibfnamefont{P.-L.} \bibnamefont{Lavertu}},
  \bibinfo{author}{\bibfnamefont{M.-A.} \bibnamefont{Marois}},
  \bibnamefont{and} \bibinfo{author}{\bibfnamefont{A.-M.~S.}
  \bibnamefont{Tremblay}}, \bibinfo{journal}{Phys. Rev. Lett.}
  \textbf{\bibinfo{volume}{94}}, \bibinfo{pages}{156404}
  (\bibinfo{year}{2005}).

\bibitem[{\citenamefont{Aichhorn and Arrigoni}(2005)}]{AA05}
\bibinfo{author}{\bibfnamefont{M.}~\bibnamefont{Aichhorn}} \bibnamefont{and}
  \bibinfo{author}{\bibfnamefont{E.}~\bibnamefont{Arrigoni}},
  \bibinfo{journal}{Europhys. Lett.} \textbf{\bibinfo{volume}{72}},
  \bibinfo{pages}{117} (\bibinfo{year}{2005}).

\bibitem[{\citenamefont{Aichhorn
  et~al.}(2006{\natexlab{a}})\citenamefont{Aichhorn, Arrigoni, Potthoff, and
  Hanke}}]{AAPH06a}
\bibinfo{author}{\bibfnamefont{M.}~\bibnamefont{Aichhorn}},
  \bibinfo{author}{\bibfnamefont{E.}~\bibnamefont{Arrigoni}},
  \bibinfo{author}{\bibfnamefont{M.}~\bibnamefont{Potthoff}}, \bibnamefont{and}
  \bibinfo{author}{\bibfnamefont{W.}~\bibnamefont{Hanke}},
  \bibinfo{journal}{Phys. Rev. B} \textbf{\bibinfo{volume}{74}},
  \bibinfo{pages}{024508} (\bibinfo{year}{2006}{\natexlab{a}}).

\bibitem[{\citenamefont{Aichhorn
  et~al.}(2006{\natexlab{b}})\citenamefont{Aichhorn, Arrigoni, Potthoff, and
  Hanke}}]{AAPH06b}
\bibinfo{author}{\bibfnamefont{M.}~\bibnamefont{Aichhorn}},
  \bibinfo{author}{\bibfnamefont{E.}~\bibnamefont{Arrigoni}},
  \bibinfo{author}{\bibfnamefont{M.}~\bibnamefont{Potthoff}}, \bibnamefont{and}
  \bibinfo{author}{\bibfnamefont{W.}~\bibnamefont{Hanke}},
  \bibinfo{journal}{Phys. Rev. B} \textbf{\bibinfo{volume}{74}},
  \bibinfo{pages}{235117} (\bibinfo{year}{2006}{\natexlab{b}}).

\bibitem[{\citenamefont{Aichhorn et~al.}(2007)\citenamefont{Aichhorn, Arrigoni,
  Potthoff, and Hanke}}]{AAPH07}
\bibinfo{author}{\bibfnamefont{M.}~\bibnamefont{Aichhorn}},
  \bibinfo{author}{\bibfnamefont{E.}~\bibnamefont{Arrigoni}},
  \bibinfo{author}{\bibfnamefont{M.}~\bibnamefont{Potthoff}}, \bibnamefont{and}
  \bibinfo{author}{\bibfnamefont{W.}~\bibnamefont{Hanke}},
  \bibinfo{journal}{Phys. Rev. B} \textbf{\bibinfo{volume}{76}},
  \bibinfo{pages}{224509} (\bibinfo{year}{2007}).

\bibitem[{\citenamefont{Kancharla et~al.}(2008)\citenamefont{Kancharla, Kyung,
  S\'{e}n\'{e}chal, Civelli, Capone, Kotliar, and Tremblay}}]{KKS+08}
\bibinfo{author}{\bibfnamefont{S.~S.} \bibnamefont{Kancharla}},
  \bibinfo{author}{\bibfnamefont{B.}~\bibnamefont{Kyung}},
  \bibinfo{author}{\bibfnamefont{D.}~\bibnamefont{S\'{e}n\'{e}chal}},
  \bibinfo{author}{\bibfnamefont{M.}~\bibnamefont{Civelli}},
  \bibinfo{author}{\bibfnamefont{M.}~\bibnamefont{Capone}},
  \bibinfo{author}{\bibfnamefont{G.}~\bibnamefont{Kotliar}}, \bibnamefont{and}
  \bibinfo{author}{\bibfnamefont{A.-M.~S.} \bibnamefont{Tremblay}},
  \bibinfo{journal}{Phys. Rev. B} \textbf{\bibinfo{volume}{77}},
  \bibinfo{pages}{184516} (\bibinfo{year}{2008}).

\bibitem[{\citenamefont{Capone and Kotliar}(2006)}]{capone:054513}
\bibinfo{author}{\bibfnamefont{M.}~\bibnamefont{Capone}} \bibnamefont{and}
  \bibinfo{author}{\bibfnamefont{G.}~\bibnamefont{Kotliar}},
  \bibinfo{journal}{Phys. Rev. B} \textbf{\bibinfo{volume}{74}},
  \bibinfo{eid}{054513} (\bibinfo{year}{2006}).

\bibitem[{\citenamefont{Potthoff}(2005)}]{Pot05}
\bibinfo{author}{\bibfnamefont{M.}~\bibnamefont{Potthoff}},
  \bibinfo{journal}{Adv. Solid State Phys.} \textbf{\bibinfo{volume}{45}},
  \bibinfo{pages}{135} (\bibinfo{year}{2005}).

\bibitem[{\citenamefont{Luttinger and Ward}(1960)}]{LW60}
\bibinfo{author}{\bibfnamefont{J.~M.} \bibnamefont{Luttinger}}
  \bibnamefont{and} \bibinfo{author}{\bibfnamefont{J.~C.} \bibnamefont{Ward}},
  \bibinfo{journal}{Phys. Rev.} \textbf{\bibinfo{volume}{118}},
  \bibinfo{pages}{1417} (\bibinfo{year}{1960}).

\bibitem[{\citenamefont{Dahnken et~al.}(2004)\citenamefont{Dahnken, Aichhorn,
  Hanke, Arrigoni, and Potthoff}}]{DAH+04}
\bibinfo{author}{\bibfnamefont{C.}~\bibnamefont{Dahnken}},
  \bibinfo{author}{\bibfnamefont{M.}~\bibnamefont{Aichhorn}},
  \bibinfo{author}{\bibfnamefont{W.}~\bibnamefont{Hanke}},
  \bibinfo{author}{\bibfnamefont{E.}~\bibnamefont{Arrigoni}}, \bibnamefont{and}
  \bibinfo{author}{\bibfnamefont{M.}~\bibnamefont{Potthoff}},
  \bibinfo{journal}{Phys. Rev. B} \textbf{\bibinfo{volume}{70}},
  \bibinfo{pages}{245110} (\bibinfo{year}{2004}).

\bibitem[{\citenamefont{Pozgajcic}(2004)}]{Poz04}
\bibinfo{author}{\bibfnamefont{K.}~\bibnamefont{Pozgajcic}},
  \bibinfo{journal}{preprint cond-mat} \textbf{\bibinfo{volume}{0407172}}
  (\bibinfo{year}{2004}).

\bibitem[{\citenamefont{Balzer et~al.}(2008)\citenamefont{Balzer, Hanke, and
  Potthoff}}]{BHP08}
\bibinfo{author}{\bibfnamefont{M.}~\bibnamefont{Balzer}},
  \bibinfo{author}{\bibfnamefont{W.}~\bibnamefont{Hanke}}, \bibnamefont{and}
  \bibinfo{author}{\bibfnamefont{M.}~\bibnamefont{Potthoff}},
  \bibinfo{journal}{Phys. Rev. B} \textbf{\bibinfo{volume}{77}},
  \bibinfo{pages}{045133} (\bibinfo{year}{2008}).

\bibitem[{\citenamefont{Balzer et~al.}(2009)\citenamefont{Balzer, Kyung,
  S\'en\'echal, Tremblay, and Potthoff}}]{BKS+09}
\bibinfo{author}{\bibfnamefont{M.}~\bibnamefont{Balzer}},
  \bibinfo{author}{\bibfnamefont{B.}~\bibnamefont{Kyung}},
  \bibinfo{author}{\bibfnamefont{D.}~\bibnamefont{S\'en\'echal}},
  \bibinfo{author}{\bibfnamefont{A.-M.~S.} \bibnamefont{Tremblay}},
  \bibnamefont{and} \bibinfo{author}{\bibfnamefont{M.}~\bibnamefont{Potthoff}},
  \bibinfo{journal}{Europhys. Lett.} \textbf{\bibinfo{volume}{85}},
  \bibinfo{pages}{17002} (\bibinfo{year}{2009}).

\bibitem[{\citenamefont{Yokoyama et~al.}(2006)\citenamefont{Yokoyama, Ogata,
  and Tanaka}}]{YOT06}
\bibinfo{author}{\bibfnamefont{H.}~\bibnamefont{Yokoyama}},
  \bibinfo{author}{\bibfnamefont{M.}~\bibnamefont{Ogata}}, \bibnamefont{and}
  \bibinfo{author}{\bibfnamefont{Y.}~\bibnamefont{Tanaka}},
  \bibinfo{journal}{J. Phys. Soc. Jpn.} \textbf{\bibinfo{volume}{75}},
  \bibinfo{pages}{114706} (\bibinfo{year}{2006}).

\end{thebibliography}

\end{document}